\documentclass{article}
\usepackage{cite}
\usepackage{amsfonts}
\usepackage{amssymb}
\usepackage{amsmath}
\usepackage{graphicx}
\usepackage{slashed}
\usepackage{setspace}

\usepackage{color}
\definecolor{grn}{rgb}{.0,0.60,0.0}
\definecolor{rd}{rgb}{.60,0.0,0.0}
\definecolor{blk}{rgb}{.0,.0,.0}

\def\tQ{\tilde{Q}}

\newcommand{\reef}[1]{(\ref{#1})}

\newcommand{\dash}{\text{-}}
\newcommand{\pdan}{{m}}

\newcommand{\cac}{{\cal C}}

\newcommand{\cn}{{\cal N}}

\newcommand{\be}{\begin{equation}}
\newcommand{\ee}{\end{equation}}

\def\be{\begin{equation}}
\def\ee{\end{equation}}
\def\bea{\begin{eqnarray}}
\def\eea{\end{eqnarray}}
\def\ba{\begin{array}}
\def\ea{\end{array}}
\def\bd{\begin{displaymath}}
\def\ed{\end{displaymath}}

\def\ie{{\it i.e.~}}

\def\a{\alpha}
\def\b{\beta}

\def\d{\delta}
\def\e{\epsilon}           
  
\def\g{\gamma}
\def\h{\eta}

\def\m{\mu}
\def\n{\nu}
  
\def\G{\Gamma}

\def\pa{\partial}                              

\def\>{\rangle} 
\def\<{\langle} 
\def\Dsl{D \hskip-.6em \raise1pt\hbox{$ / $ } }
\def\to{\rightarrow}

\def\pa{\partial}

\def\lab{\label}

\newcommand{\lra}{\leftrightarrow}

\def\tQ{\tilde{Q}}

\begin{document}

\setstretch{1.05}

\begin{titlepage}
\begin{flushright}
MIT-CTP-4131 \\
MCTP 10-10 \\
PUPT-2334 \\
\end{flushright}
\vspace{1cm}

\begin{center}
{\Large\bf A simple approach to counterterms in N=8 supergravity}  \\
\vspace{1cm}
{\bf Henriette Elvang${}^{a,b}$,
Daniel Z.~Freedman${}^{c,d}$, Michael Kiermaier$^{e}$} \\
\vspace{0.7cm}
{${}^{a}${\it Michigan Center for Theoretical Physics}\\
{\it Randall Laboratory of Physics}\\
{\it University of Michigan}\\
{\it 450 Church St, Ann Arbor, MI 48109, USA}}\\[5mm]
{${}^{b}${\it School of Natural Sciences}\\
         {\it Institute for Advanced Study}\\
         {\it Princeton, NJ 08540, USA}}\\[5mm]
{${}^{c}${\it Center for Theoretical Physics}}\\
{${}^{d}${\it Department of Mathematics}}\\
         {\it Massachusetts Institute of Technology}\\
         {\it 77 Massachusetts Avenue}\\
         {\it Cambridge, MA 02139, USA}\\[5mm]
{${}^{e}${\it Joseph Henry Laboratories}\\
{\it Princeton University}\\
{\it Princeton, NJ 08544, USA}}\\[5mm]
{\small \tt  elvang@umich.edu,
 dzf@math.mit.edu, mkiermai@princeton.edu}
\end{center}
\vskip .3truecm
\begin{abstract}
We present a simple systematic  method to study candidate counterterms in $\cn=8$
supergravity.  Complicated details of the counterterm operators are avoided because
we work  with the on-shell matrix elements they produce.  All $n$-point matrix elements of an  independent SUSY invariant operator of the form $D^{2k} R^n +\dots$ must be local and satisfy SUSY Ward identities.  These are strong constraints, and
we test directly whether or not matrix elements with these properties can be constructed. If not, then the operator does not have a  supersymmetrization, and it is excluded as a potential counterterm.  For $n>4$, we find that $R^n$, $D^{2} R^n$, $D^{4} R^n$, and $D^{6} R^n$ are excluded as counterterms of MHV amplitudes, while only $R^n$ and $D^{2} R^n$ are excluded at the  NMHV level.
As a consequence, for loop order $L<7$, there are no independent $D^{2k}R^n$ counterterms with $n>4$. If an operator is not ruled out, our method constructs an explicit superamplitude  for its matrix elements. This is done for the 7-loop  $D^4 R^6$ operator at the NMHV level and in other cases.  We also initiate the study of counterterms without leading pure-graviton matrix elements, which can occur beyond the MHV level. The landscape of excluded/allowed candidate counterterms is summarized in a colorful chart.

\end{abstract}
\end{titlepage}

\setstretch{0.8}
\tableofcontents
\setstretch{1.05}

\newpage
\newpage
\setcounter{equation}{0}
\section{Introduction and Summary of Results}

At which loop order does the first UV divergence occur in $\cn = 8$ supergravity in four dimensions? Could the theory possibly be perturbatively finite? These old enticing questions have recently received renewed attention.
Impressive calculations in field theory \cite{Bern:2007hh,Bern:2008pv,Bern:2009kf,Bern:2009kd} --- with unexpected cancellations \cite{Bern:2007hh,Bern:2008pv,Bern:2009kf,Bern:2009kd,BjerrumBohr:2006yw,Bern:2007xj,BjerrumBohr:2008vc,BjerrumBohr:2008ji,ArkaniHamed:2008gz} even for dimensions $D>4$ --- have explicitly demonstrated that  four-point 3- and 4-loop amplitudes are finite.
In addition both superspace formulations \cite{
Howe:1980th,Kallosh:1980fi,Howe:2002ui,Drummond:2003ex,Stelle:2007zz,Bossard:2009sy,Bossard:2009mn,Stelle:2009zz} and string theory methods \cite{
Berkovits:2006vc,Green:2006gt,Green:2006yu,Berkovits:2009aw,Green:2010wi,Green:2010sp} have been explored to rule out some operators as potential counterterms and identify others as the likely first divergence. An important  goal of such studies is  the 
classification of operators that are viable candidate counterterms of $\cn=8$ supergravity.

The purpose of this paper is to introduce a simple systematic 
method to study the supersymmetrization 
of local operators as a test of whether they can be candidate counterterms in $\cn=8$ supergravity. Consider the operator $D^{2k} R^n$,
which denotes an unspecified Lorentz invariant contraction of $2k$ covariant derivatives and $n$ Riemann tensors.
To be a candidate counterterm, it must have an $\cn=8$ supersymmetrization, which we denote schematically by ``$D^{2k} R^n+\ldots$". This could appear at loop level $L=n+k-1$. Rather than examine the operator
directly, we study the $n$-point on-shell matrix elements it would generate.
SUSY requires that these on-shell matrix elements satisfy SUSY Ward identities. Furthermore, locality implies that they have no poles in any momentum channels.
If the combined constraints of locality and SUSY cannot be satisfied for any matrix elements of the putative term $D^{2k} R^n+\ldots$, then such a
supersymmetrization does not exist. In that case we can rule out $D^{2k} R^n+\ldots$ as a candidate counterterm. On the other hand, in cases where SUSY Ward identities and locality are compatible, our method constructs the matrix elements explicitly. The SUSY Ward identities test supersymmetry at the linearized level, and the matrix elements
correspond to a linearized supersymmetrization of the operator.

The index contractions  of an operator $D^{2k}R^n+ \dots $ can be organized according to the N$^K$MHV classification of its $n$-point matrix elements. This is
 possible because on-shell the Riemann tensor $R_{\mu\nu\rho\sigma}$  splits  into a totally symmetric 4th rank spinor $R_{\a\b\g\d}$ and its conjugate $\bar{R}_{\dot{\a}\dot{\b}\dot{\g}\dot{\d}}$, which communicate to gravitons of opposite helicity. Terms in
$D^{2k}R^n$ with 2 factors of $R$ and $(n-2)$ factors of $\bar{R}$ contribute to the MHV graviton matrix element while $R^3\,\bar{R}^{n-3}$ is the NMHV part and so on.  This separation persists in the SUSY completion, because amplitudes in each N$^K$MHV sector satisfy Ward identities independently.

We examine the MHV and NMHV matrix elements of each operator separately, but never need to concern ourselves with its specific index contractions.
Indeed the only input needed is the mass dimension (for $D^{2k} R^n+\ldots$ it is $2(k+n)$). This information is combined with little group scaling requirements \cite{Witten:2003nn} to study the possible local $n$-particle matrix elements
 that the operator generates. This allows us to construct the most general expressions for the local ``basis matrix elements" needed to determine the MHV and NMHV generating functions (also called  `superamplitudes'). At the NMHV level we use the manifestly supersymmetric ``basis expansion" for  superamplitudes 
derived in \cite{efk4}. Individual matrix elements (projected out from these superamplitudes by Grassmann differentiation) are linear combinations of the basis matrix elements, and they automatically satisfy the SUSY Ward identities.  However,  the input of local \emph{basis} matrix elements is not necessarily sufficient to make \emph{all} matrix elements local. We utilize a complex shift to identify
cases where locality fails. The failure of locality means that supersymmetrization of the operator is ruled out. When locality succeeds, the method produces the explicit permutation symmetric superamplitude that 
generates the matrix elements of the linear supersymmetrization of the operator.

\begin{figure}
\includegraphics[width=15.3cm]{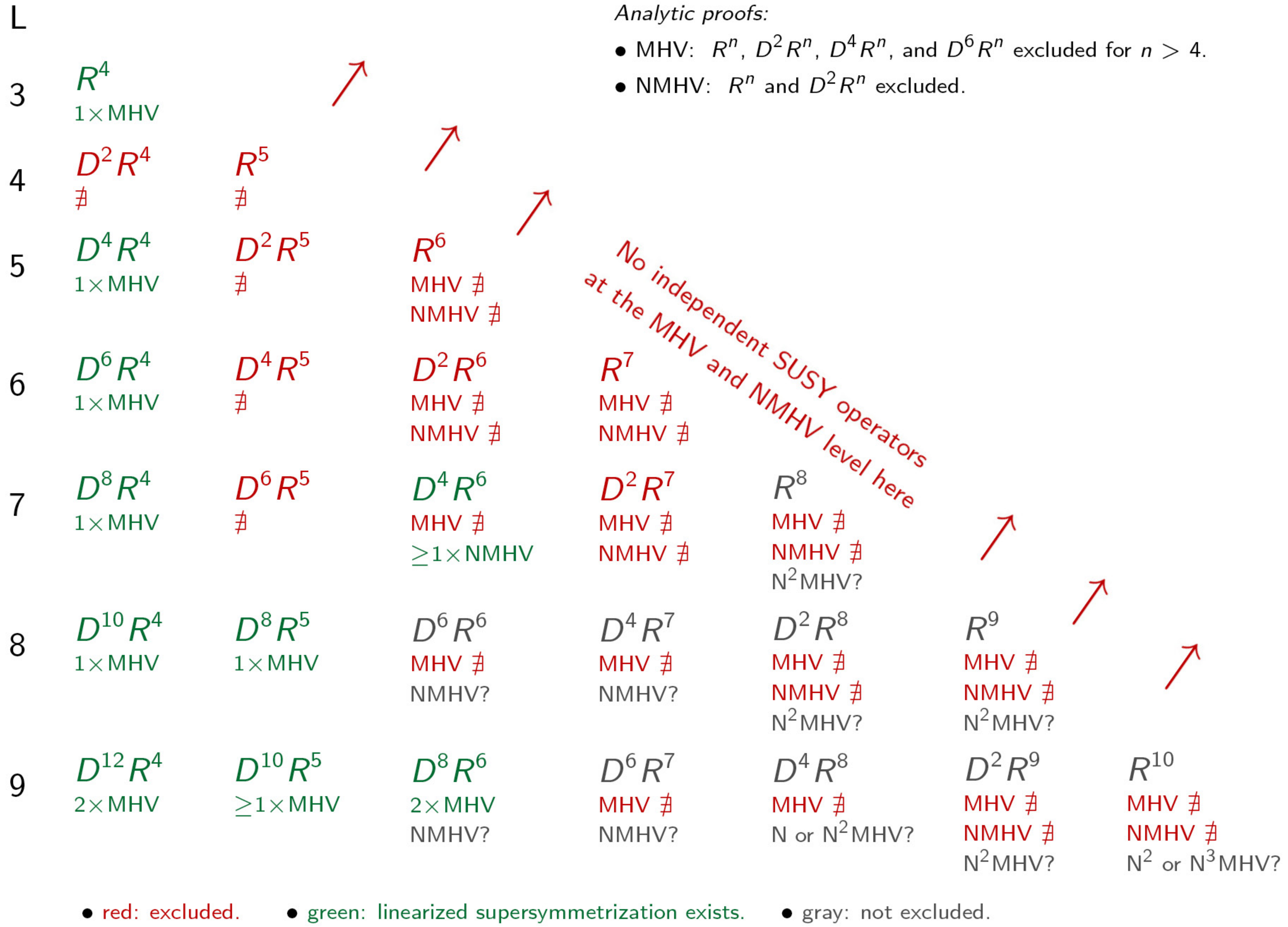}
\caption{{\small Results for candidate counterterms in $\cn=8$ supergravity, organized by loop order $L$ and $n$-point level of their leading matrix elements. The color indicates whether a
linearized
supersymmetrization
of the $D^{2k}R^n$ operator under consideration exists (green), is excluded (red), or is unknown (gray). Beyond the MHV level, there could be SUSY operators without a leading pure-graviton contribution.
These could also ``live'' above the $R^n$ diagonal in this diagram. In section \ref{secnograv}, we rule out such operators at the NMHV level above the $D^4 R^n$ line.
}}
\label{tabsummary}
\end{figure}

At the MHV level, the method allows us to rule out the existence of independent supersymmetric completions of all operators of the form $R^n$, $D^2R^n$, $D^4R^n$, and $D^6R^n$, for any $n\geq5$. On the other hand, linearized supersymmetrizations
of $D^8R^n$ MHV operators exist for all $n \ge 4$, as we show through an explicit construction. At the NMHV level, we rule out supersymmetrizations
of
all $R^n$ and $D^2R^n$ operators. However, we do not rule out $D^4R^n$ operators at NMHV level. In fact, we explicitly construct an NMHV superamplitude satisfying all requirements for a
linearized
supersymmetric completion of $D^4R^6$. An overview of our results is provided in Fig.~\ref{tabsummary}. We discuss these results further in section \ref{secDisc}.

In the remainder of this section we outline the consequences for candidate counterterms. Counterterms for the possible UV divergences of $\cn=8$ supergravity must be  local, supersymmetric operators with $SU(8)_R$ symmetry.\footnote{It was shown in \cite{Marcus:1985yy,diVecchia:1984jh} that the $SU(8)$ R-symmetry is non-anomalous.}  Our method tests whether operators can have a linearized supersymmetrization:
if an operator fails the test, it can be excluded as a candidate 
 counterterm for the first divergence in $\cn=8$ supergravity.
On the other hand, if an operator passes  the  tests we cannot state whether it actually appears in the perturbation expansion.  Further information --- perhaps  from explicit loop calculations or additional  symmetries such as $E_{7,7}(\mathbb{R})$ --- will be required to decide this question.

Previously most analyses have focused on potential divergences in 4-particle amplitudes (see, however, \cite{Drummond:2003ex}). With our new approach we address two types of questions:
\begin{enumerate}
\item  Suppose it is shown that the 4-point $L$-loop amplitude is UV finite. Does this suffice to rule out UV divergences in all $L$-loop amplitudes? In gravity, the naive power counting of higher-point amplitudes is the same as for the 4-point amplitudes, since every bosonic vertex  in the classical action is quadratic in momenta. Hence it requires a separate analysis to establish finiteness for all $n$-point amplitudes at a given loop order. For example, at the 5-loop level, finiteness of the 4-point amplitude would eliminate $D^4 R^4$ as a counterterm. But without further information one cannot exclude higher-point amplitudes whose divergences generate independent counterterms such as $D^2 R^5$ and $R^6$. We provide such an analysis.

In sections \ref{secMHV} and \ref{secNMHV}, we show that  for $L$-loop $n$-point amplitudes with $n > 4$ no independent SUSY candidate counterterms exist for $n> L-3$ at the MHV level\footnote{This bound was also noted in \cite{Kalloshtalk}.} and for $n>L-1$ at the NMHV level. 

\item Can there be \emph{independent}
SUSY candidate counterterms without a leading pure graviton contribution? Clearly the operators listed in Fig.~\ref{tabsummary} all give rise to matrix elements whose external states are all gravitons. So the question is whether there could be  other 
independent counter\-terms,for example 
``above the  $R^n$ diagonal'' in the chart. Such operators could not contribute at the MHV level, since all such amplitudes are proportional to the pure-graviton amplitude.
But  beyond the MHV level this is more subtle. We explain why and address this point in section \ref{secnograv}. In particular, we show that no  SUSY operators ``above the $R^n$ diagonal'' in Fig.~\ref{tabsummary}  exist at the NMHV level. Beyond the NMHV level, such independent SUSY invariants may exist, but we propose a lower bound on their mass dimension. If the conjectured bound is true, operators ``above the $R^n$ diagonal'' cannot appear at loop order $L<7$.
\end{enumerate}

It is a  direct consequence of our results that no supersymmetrizations of $D^{2k}R^n$ with $n>4$ exist for loop levels $L<7$. If the conjectured  bound on the mass dimension of beyond-NMHV operators holds, then the only independent counterterms for $L<7$ are the supersymmetrizations of the 4-graviton operators $D^{2k} R^4$. 
Earlier work \cite{Drummond:2003ex} pointed out that no higher-point supersymmetric operators are constructible as subsuperspace integrals in harmonic superspace.

At the 7-loop level, we rule out all MHV candidate counterterms, except $D^8 R^4$. 
At the NMHV level our analysis rules out $D^2 R^7$ and $R^8$. However, we find that an independent linearized supersymmetrization of $D^4 R^6$ exists. Thus it is a candidate counterterm for a 7-loop divergence in 6-point NMHV amplitudes. It would be interesting to extend our analysis beyond the MHV and NMHV levels. This would facilitate an analysis of the N$^2$MHV matrix elements of the 7-loop $R^8$ operator and test our   conjecture  that for $L<7$
no SUSY operators exist above the diagonal in Fig.~\ref{tabsummary}.

\noindent {\sl Note:}
The work presented here has some overlap with an approach by Kallosh and
Rube \cite{Kalloshtalk,Kallosh:2009db}, but there are important differences in
both the methods and results.
For example, motivated by a light-cone gauge approach, refs.~\cite{Kalloshtalk,Kallosh:2009db} impose a certain locality requirement on the superamplitude. We, more conservatively, only impose locality on the matrix elements it produces. It is the latter property that is directly implied if a gauge invariant local counterterm $D^{2k} R^n +\dots$ exists.


\setcounter{equation}{0}
\section{A matrix element approach to counterterms}

We discuss counterterms in section \ref{secLCTs} before introducing the matrix element method in section  \ref{secMEs}.
We explain in section \ref{sec:perm} how full permutation symmetry of the superamplitude allows us to devise a simple test of locality.

\subsection{Local counterterms}
\lab{secLCTs}

We begin by describing some facts about candidate counterterms in the
perturbative expansion of $\cn=8$ supergravity.
 Naive power counting indicates that loop amplitudes in gravity are divergent.
 Calculations are generally done using dimensional regularization. In this framework one is concerned with possible $1/\e$ poles in on-shell amplitudes. If such poles occur  they can be canceled by local counterterms constructed from the classical fields of the theory. Since the gravitational coupling is dimensionful, the loop order determines the mass dimension of the counterterm. For example, the dimension $2(n+k)$ operator $D^{2k}R^n$ is a candidate counterterm for $n$-graviton amplitudes at loop order $L=n+k-1$. The graviton operator $D^{2k} R^n$ can only appear as a 
counterterm\footnote{There appears to be very little information available about the specific component form of the superspace counter\-terms discussed in the literature. The component expansions of $R^4$ and $F^4$ have been constructed in \cite{Peeters:2000qj} through quadratic order in fermions in 10
dimensions. Concerning $R^4$ in 4 dimensions, see \cite{Gross:1986iv}, and in 11 dimensions \cite{SD}.} if it can be supersymmetrized.

Our discussion of counterterms pertains directly to the lowest-point 
 amplitude at the lowest loop level 
 $L$ for which divergences actually occur in the perturbative S-matrix of $\cn=8$ supergravity.  We emphasize this point because
the nonlinear supersymmetric completion of $D^{2k}R^n$ may require terms that are higher order in the Riemann tensor, such as $D^{2k-2}R^{n+1}$, in addition to terms involving other fields of the $\cn=8$ multiplet. A relevant example is the $\cn=4$ SYM operator $D^2F^4 + F^5$, which appears in the $\alpha'$-expansion of open string tree amplitudes \cite{Drummond:2003ex}.
The 4-point local matrix elements of this operator satisfy the $\cn=4$ SUSY Ward identities, but the individual local 5-point matrix elements
 of the $F^5$ contribution do not. This is consistent because the
 5-point matrix
 elements of the complete operator are actually \emph{not} local. They contain contributions from pole diagrams involving the 4-point vertex in $D^2F^4$ and the 3-point vertex of the classical action in addition to the 5-point local terms  from both $D^2F^4$ and $F^5$. In $\cn =8$ supergravity, it is possible that the completion of a candidate counterterm $D^4R^4$ contains such \emph{dependent} $D^2R^5$ and $R^6$ terms.\footnote{Stieberger \cite{Stieberger:2009rr} has studied (non-local) 5- and
6-point matrix elements of $D^{2k}R^4$ from the expansion of the string amplitude.}
 The $D^2R^5$ and $R^6$ terms
ruled out by our methods are \emph{independent} structures
 that satisfy linearized $\cn=8$ SUSY. Similarly, $D^{10}R^4$ may contain dependent $D^8R^5$, while the admissible $D^8R^5$ listed in
Fig.~\ref{tabsummary} is an independent linearized invariant. When we consider the counterterm of the first divergence in the theory, this issue of lower-point operators does not arise.

\subsection{The method}
\lab{secMEs}

In our approach we focus directly on general local
$n$-point matrix elements of $D^{2k}R^n+\ldots$\,, with arbitrary sets of particles of the $\cn=8$ theory,  namely  $m_n(p_i, h_i),~i=1,\dots, n$, where $p_i$ and $h_i$ denote the 4-momentum and helicity of the $i$th particle. We use the
spinor-helicity formalism in which momenta
and polarizations
are encoded in spinors $|i\>$ and $|i]$. For $n \ge 4$, 
\emph{local} matrix elements $m_n$ are polynomials in angle and square brackets; uncanceled brackets $\<ij\>$ or $[k l]$ cannot occur in the denominator of any matrix element. Details of this argument are given in Appendix \ref{appA}.

``Little group scaling''  \cite{Witten:2003nn} requires
\bea
  m_n(\dots, t_i |i\> , t_i^{-1} |i], \dots) = t_i^{-2h_i}\, m_n(\dots, |i\> , |i], \dots )
\eea
for each particle $i$. This determines that the number of factors $a_i$ of the angle spinor $|i\>$ minus the number of factors $s_i$ of the square spinor $|i]$ for each particle in $m_n$ is 
$a_i - s_i = -2h_i$.  Helicity scaling weights and overall
dimension provide important constraints on the structure of the local matrix elements obtained from a candidate counterterm.

In some cases, these constraints determine the matrix element completely. A well known example is the $R^4$ 4-graviton matrix element $m_4(1^-,2^-,3^+,4^+)$, which has overall dimension 8 and
 spinor content $|1\>^4$, $|2\>^4$, $|3]^4$, $|4]^4$\,, which follows from the
helicity weights $-2h_i = +4,\,+4,\, -4,\, -4$\,.   The only local expression that carries the
correct dimension and weights is
\be
  m_4(1^-,2^-,3^+,4^+)= \<12\>^4[34]^4\, .
\ee
This form also appears, for example,  in \cite{Kallosh:2008mq}.
The better known \cite{Gross:1986iv}  form
\be
\lab{m4}
  m_4(1^-,2^-,3^+,4^+)= s\, t\, u\, M_4^{{\rm tree}}
  = s\, t\, u\; \frac{\<12\>^8[12]}{\<12\>\<13\>\<14\>\<23\>\<24\>\<34\>^2}
\ee
appears to be
 non-polynomial in spinor brackets,
but it can be shown to be equal to the first expression using momentum conservation.

The requirement that a candidate field theory counterterm is invariant under $\cn=8$ supersymmetry translates into the statement that the matrix elements obtained from them satisfy SUSY Ward identities.
 To facilitate the Ward identity analysis, we work with superamplitudes,
containing Grassmann bookkeeping variables $\h_{ia},~ a=1,\dots, 8$.

The MHV sector is particularly simple because there each SUSY counterterm has only one independent matrix element.
We are free to take this to be the $n$-graviton matrix element $m_n(--+\dots+)$ (which is shorthand for $m_n(1^-,2^-,3^+,\dots,n^+)$).
The
 $n$-particle superamplitude representing the counterterm is a 16th order Grassmann polynomial, which then
 takes the form  \cite{Witten:2003nn} 
\be \lab{mhvsamp}
\cac^{\rm MHV}_n =
\d^{(16)}\Bigl(\sum_i | i\> \eta_{ai}\Bigr) \frac{m_n(--+\dots+)}{\<12\>^8}\,.
\ee
Note that the matrix element $m_n(--+\dots+)$ must be bose symmetric under exchange of helicity spinors for the two negative helicity particles and for any pair
of positive helicity particles.   The superamplitude must have full $S_n$ permutation symmetry, and so must the ratio 
$m_n(--+\dots+)/\<12\>^8$\,. 
Specific matrix elements for any MHV process in the theory are obtained by applying Grassmann derivatives of order 16 as described in \cite{BEF}.

The analysis extends to the NMHV level  where we study potential
 $D^{2k} R^n$ counterterms for $n \ge 6$.  The NMHV superamplitude is a Grassmann polynomial of order 24.
We use the  manifestly SUSY and R-symmetry invariant
representations recently derived in \cite{efk4}, which express  superamplitudes as  sums
 that involve several independent basis matrix elements.  For each of these basis elements one needs a local expression with the correct little group scaling properties.

Our general strategy to study linearized counterterms is a two step procedure:
\begin{enumerate}
\item  For each basis matrix element, write down the most general polynomial in spinor brackets consistent
with
 particle exchange symmetries, helicity weight $-2h_i$ for each particle and fixed total mass dimension.
 Use these general local basis matrix elements to construct
 a general  ansatz for the superamplitude.
\item In the second step, we either \emph{exclude} a counterterm  or \emph{construct} its most general matrix elements compatible with SUSY:
    \begin{itemize}
    \item
 {\bf To rule out the existence of a counterterm:}\\
 Show that at least one matrix element computed from 
 the  general superamplitude ansatz is non-local. To show this, we demonstrate that the matrix element has a pole under a complex on-shell deformation of the external momenta.
 \item
  {\bf To construct a candidate counterterm:}\\
 Verify that a \emph{fully $S_n$ permutation-symmetric} superamplitude exists within the ansatz. It will be shown below in section \ref{sec:perm} that permutation symmetry together with locality of the basis matrix elements guarantees that all matrix elements produced are local.
 \end{itemize}
\end{enumerate}

Since the SUSY Ward identities
 used in the superamplitude construction of~\cite{efk4} are those obtained from the lowest order transformation rules of the fields of $\cn =8$ supergravity,  any independent candidate counterterm that satisfies our requirements is established only at the linearized level.

\subsection{From permutation symmetry to locality}
\lab{sec:perm}

In this section, we show that \emph{any superamplitude with local basis matrix elements and full permutation symmetry produces local matrix elements for any process.} 

We first consider the special case of MHV superamplitudes.
Any pole in a matrix element computed from the MHV superamplitude $\cac^{\rm MHV}_n$ in~(\ref{mhvsamp}) with local basis matrix element $m_n(--+\dots+)$ can only arise from the factor 1/$\<12\>^8$. However, if a matrix element computed from a \emph{permutation symmetric} $\cac^{\rm MHV}_n$ had a pole in $\<12\>$, then $\cac^{\rm MHV}_n$ would also produce a permuted matrix element with a pole in, say, $\<34\>$. But poles in $\<34\>$ are manifestly absent in $\cac^{\rm MHV}_n$. So we conclude that any matrix element computed from a permutation-symmetric MHV superamplitude $\cac^{\rm MHV}_n$ with local basis matrix element $m_n(--+\dots+)$ is local. 

This argument immediately generalizes beyond the MHV level. To see this, consider the  manifestly SUSY and R-invariant $n$-point superamplitude of~\cite{efk4} at the N$^K$MHV level. For the current argument, most details of the construction of \cite{efk4} are not needed. We simply note that the N$^K$MHV superamplitude requires several basis
matrix elements $m^{(j)}$ and takes the schematic form\footnote{Readers can look ahead to eqs.~\reef{SGNMHV} and \reef{xpoly8} for the precise form of the 6-point NMHV superamplitude.}
\begin{equation} \label{CNKMHV}
    \cac^\text{N$^K$MHV}_n~=~\sum_j m^{(j)}X_{(j)}\,.
\end{equation}
The $X_{(j)}$ are SUSY and $SU(8)_R$ invariant Grassmann polynomials of order $8(K+2)$. Each of them includes the factor $\d^{(16)}(\sum_i | i\> \eta_{ai})$,
 which also appeared in \reef{mhvsamp}.
With local basis matrix elements inserted, this superamplitude has the structure
\begin{equation}
    \cac^\text{N$^K$MHV}_n~=~\frac{\text{manifestly local}}{[n\dash 3,n\dash 2]^{8K}\<n\dash 1,n\>^8}\,.
\end{equation}
For any choice of local basis matrix elements, this superamplitude can only have poles in $[n\dash 3,n\dash 2]$ and $\<n\dash 1,n\>$. It cannot give rise to poles in other angle and square brackets, such as $[12]$ or $\<34\>$. Therefore, if the superamplitude is permutation invariant, all matrix elements obtained from it cannot have poles in any spinor brackets. They must be local.

Verifying permutation symmetry is thus a crucial step in the construction of candidate counter\-terms. Let us explain how permutation symmetry  can be checked in practice.
Consider two renditions of a
 superamplitude, $\mathcal{F}$ and $\tilde{\mathcal{F}}$, which differ by
a permutation of the momentum labels in their basis elements and
$X$-polynomials. If the  superamplitude is permutation symmetric then
the two renditions are equal, so we write
\be  \lab{F1F2}
\sum_j m^{(j)} X_{(j)} =
\mathcal{F} =\tilde{\mathcal{F}}
= \sum_j \tilde{m}^{(j)}  \tilde{X}_{(j)}\, .
\ee
If the basis elements $\tilde{m}^{(j)}$ of $\tilde{\mathcal{F}}$ are projected out correctly from $\mathcal{F}$, then the equality  \reef{F1F2} holds because the basis matrix elements uniquely determine the 
 superamplitude.  To test \reef{F1F2} we
therefore apply the Grassmann derivative
 $\tilde{D}^{(i)}$ appropriate to the particle
states of a particular basis element $\tilde{m}^{(i)}$ and obtain
\be
 \sum _j a_{ij} m^{(j)} = \tilde{D}^{(i)} \mathcal{F} = \tilde{D}^{(i)} \tilde{\mathcal{F}} = \tilde{m}^{(i)}.
\ee
Generically all basis elements
of
 $\mathcal{F}$ contribute to the sum on the left side. However, if the sum
reproduces $\tilde{m}^{(i)}$ for all
$i$, then $\mathcal{F}$ and $\tilde{\mathcal{F}}$ are consistent and therefore equal.
It is a straightforward and systematic procedure to verify \reef{F1F2} for all basis elements. For full permutation symmetry, one has to repeat the above check for a set of permutations that generates the group $S_n$.

On the other hand, checking locality directly is much more difficult. In principle one would have to explicitly compute the matrix elements for each possible
process (not just permuted basis processes) and verify locality. The above
argument allows us to circumvent this lengthy procedure. This will be
important in the explicit construction of the $D^4R^6$ candidate
counterterm in section \ref{secD4R6}.


\setcounter{equation}{0}
\section{Candidate MHV counterterms}\label{secMHV}
\subsection{The $R^4$ counterterm}
\lab{secR4}

It is well known that the operator
 $R^4+\ldots$  is compatible with linearized $\cn =8$ supersymmetry \cite{Deser:1977nt,Kallosh:1980fi}. However, 3-loop calculations \cite{Bern:2007hh} show that it does not actually appear in the perturbative S-matrix.  To demonstrate how our methods work in the simplest context, we show that $R^4$ passes the tests required of a candidate counterterm (although it is a candidate which has lost the election).
The superamplitude of this counterterm is obtained by inserting the basis matrix element $m_4(--++)= \<12\>^4[34]^4$ into~(\ref{mhvsamp}),
\begin{equation}\label{R4gf}
    \cac_{R^4}^{\rm MHV}~=~
    \d^{(16)}\Bigl(\sum_i | i\> \eta_{ai}\Bigr) \frac{[34]^4}{\<12\>^4}\,. 
\end{equation}
Note that the ratio $[34]^4/\<12\>^4$  has full $S_4$ permutation symmetry, which can be easily verified using momentum conservation
\be \lab{momcons}
\<y x\>[x z] ~=~ -\!\!\sum_{i \ne x,y,z} \<y i\>[i z]\,.
\ee
All individual matrix elements  obtained by differentiation of~(\ref{R4gf}) must be local, and thus expressible as polynomials in spinor brackets. This follows immediately from the permutation symmetry of~(\ref{R4gf}), as we argued in section~\ref{sec:perm}, but we now illustrate this property explicitly.
Consider, for example, the 4-graviton matrix element $m_4(++--)$ with negative helicities on lines 3 and 4. We find
\be \lab{R4test}
m_4(++--)~=~  \biggl[\,\prod_{a=1}^8 \frac{\pa}{\pa\h_{3a}}\frac{\pa}{\pa\h_{4a}}\biggr]\,\cac_{R^4}^{\rm MHV}~=~\<34\>^8\times \frac{[34]^4}{\<12\>^4}
~=~ \<34\>^4[12]^4\,.
\ee
The singular denominator cancels in the last step because $s_{12} = s_{34}$. This form is clearly local and correctly reproduces $m_4(--++)$ with permuted momenta $(1,2) \lra (3,4)$. Using momentum conservation in the more general form  \reef{momcons} one can show that all matrix elements computed from~$\cac_{R^4}^{\rm MHV}$ are local, just as the general argument of section~\ref{sec:perm} guaranteed.

\subsection{No $D^{2}R^4$ counterterm}

It is known that $D^2 R^4$ is not an allowed candidate counterterm in $\cn=8$ supergravity \cite{Drummond:2003ex}. Let us show how this fact follows from our method. To start we consider the possible local expressions of dimension 10 for the matrix element $m_4(--++)$. The helicity weights $|1\>^4,\, |2\>^4,\, |3]^4,\, |4]^4$ account for 8 powers of  momenta, so we have to include two arbitrary spinor
pairs $|q\>[q|$ and $|q'\>[q'|$. We let $q, q'$ run over $1,2,3,4$ and in each case construct all possible spinor contractions consistent with bose symmetry $1 \lra 2$ and $3 \lra 4$. Schouten relations and momentum conservation relate the five terms found in this way, and there is only one independent local expression. Thus up to an overall constant,
$m_4(--++) =  \<12\>^5 [12] [34]^4 = s\, ( s\, t\, u\, M_4^\text{tree})$.
In the previous section, we showed that $(s\, t\, u\, M_4^\text{tree})/\<12\>^8$ is local and fully symmetric by itself.
The  superamplitude
$\delta^{(16)}(\dots) m_4(--++)/\<12\>^8$ 
must be fully symmetric in order to correctly reproduce permuted matrix elements. Symmetrizing our expression above gives a factor of $s+t+u = 0$. Thus kinematics excludes $D^2 R^4$ as a candidate counterterm.


\subsection{No  $R^n$ MHV counterterms for any $n \ge 5$}

We consider a potential counterterm  $R^n + \ldots$
of dimension $2n$ and study its $n$-point MHV matrix elements. Helicity scalings require a net count of the spinors $|1\>^4,\, |2\>^4$ and $|i]^4,\, i=3,\dots,n$. With these weights, the total dimension $2n$ is saturated, so the basis matrix element in \reef{mhvsamp} must take the form
\be \lab{Rnmhv}
m_n(--+\dots+) ~=~  \<12\>^4 f_n(|3],|4],\ldots |n])\,.
\ee
The function $f_n$ is an order $2n-4$ polynomial in square brackets, and
depends only on square spinors $|i]$ for positive helicity gravitons, \ie $i \ge 3$.
The structure \reef{Rnmhv} was also found in
[\citen{Kallosh:2008mq,Kallosh:2008ru},\,\citen{Kallosh:2009db}]. 

The basis matrix element is local, but we must test whether all other matrix elements obtained by differentiation of
\reef{mhvsamp} are also local.   We  project out 
the $n$-graviton matrix element with the negative helicity gravitons on lines 3 and 4:
\be \lab{Rn34}
m_n(++ -- + \dots +) ~=~  \frac{\<34\>^8}{\<12\>^4}\,f_n(|3],|4],\ldots |n])\,.
\ee
We now show that the
non-locality
in $\<12\>$ does not cancel for $n\geq 5$.
To do this we introduce a complex variable $z$ and
evaluate \reef{Rn34}  using the shifted spinors
\be  \lab{shift}
| i \>  \to |\hat{i}\> ~=~ | i \> + z c_i |\xi\>\,, \qquad\qquad  i =1,2,5\,,\qquad \sum_i c_i |i] =0\,,
\ee
and  all other angle spinors and all square spinors unshifted.\footnote{This type of ``holomorphic'' shift was proposed in \cite{risager}, and used in \cite{EFK1,EFK2}, to facilitate the proof of the CSW recursion relations \cite{csw}.}
  The quantity $|\xi\>$ is an arbitrary reference spinor. The shift affects only the denominator  in \reef{Rn34}, so the right-hand side has an uncanceled 4th order pole in $z$. This is inconsistent with the requirement that $m_n(++ -- + \dots +)$ is local. Hence this rules out any $R^n$  MHV counterterms for $n\ge 5$.

For $n=4$, we found above that the apparent pole
in $\<12\>$ cancels after an appropriate use of momentum conservation, rendering $m_4(++--)$ local.
It is instructive to see why the shift argument above breaks down in this case. Under a holomorphic shift, the shifted spinors must satisfy the momentum conservation condition $\sum_i c_i |i] =0$, so this type of shift cannot be implemented for 
less than 3 lines. It is impossible to construct a holomorphic shift that shifts the denominator, but leaves invariant the numerator of $m_4(++--)$ in~(\ref{R4test}). The shift argument is therefore only valid for $n\ge 5$.


\subsection{No  $D^{2}R^n$, $D^{4}R^n$, or $D^{6}R^n$ MHV counterterms for $n\ge 5$}

We now apply the same shift argument to exclude counterterms of the form $D^{2k} R^n$ for $n\geq5$
and $k<4$. Such dimension $2(k+n)$ counterterms could arise from ultraviolet divergences of $n$-particle supergravity amplitudes at loop order $L= n+k-1$.
Scaling symmetries require that  the $n$-graviton matrix element $m_n(--++\dots+)$ of this counterterm is a local polynomial with the net spinor count $|1\>^4$, $|2\>^4$ and $|i]^4$ for $i=3,\dots,n$.
These spinors comprise $2n$ powers of momenta, so we must include $2k$ pairs $|q_i\> [q_i|$ with zero net scaling to match the total dimension. Thus  $m_n(--++\dots+)$ consists of $4+k$ angle brackets and $2(n-2)+k$ square brackets.

If acceptable as a counterterm, the superamplitude
\be \label{MHVcase}
   \cac^{\rm MHV}_{D^{2k} R^n}~=~ \d^{(16)}\Bigl(\sum_i | i\> \eta_{ai}\Bigr)~\frac{m_n(--++\dots+)}{\<12\>^8}
\ee
must produce local matrix elements for any choice of external particles. To test this, we examine the permuted $n$-graviton matrix element $m_n(++--+\dots+)$.   The appropriate $\eta$ derivative applied to  \reef{MHVcase} produces the matrix element
\be \lab{MHVtest}
  m_n(++--+\dots+)= \frac{\<34\>^8}{\<12\>^8}\,m_n(--++\dots+) \, .
\ee
Let us again act with the shift~(\ref{shift}) on the angle spinors $|1\>$, $|2\>$ and $|5\>$ in this matrix element.
The factor $\<34\>^8/\<\hat{1}\hat{2}\>^8$ has an 8th order pole in the $z$-plane. The matrix element $m_n(--++\dots+)$ on the
 right-hand side of \reef{MHVtest} contains  $4+k$ angle brackets.
Thus, for $k<4$ the numerator is at most a 7th  order  polynomial in $z$. This is insufficient to cancel the 8th order pole in the denominator, so the permuted matrix element  $m_n(++--+\dots+)$ cannot be local.
We conclude that the potential MHV counterterm $D^{2k} R^n$, $n\ge 5$, cannot exist when $k<4$,  or, equivalently, at loop order $n < L-3$. 
(This bound was also given in \cite{Kalloshtalk}.)

\subsection{Candidate MHV counterterms $D^{2k} R^n$ for $k \ge 4$}

We now demonstrate that $D^{2k}R^n$ is not excluded for  $k\ge4$. Specifically, we show that there are local matrix elements that satisfy all our constraints. For the basis element $m_n(--++\dots+)$, scaling properties and overall dimension
 can now be satisfied, for example,
by polynomials of the form
$\<12\>^8 (\text{rest})$, where the ``rest'' is still polynomial in angle and square spinors.
Since the pole in the denominator of \reef{MHVtest} is directly canceled by the factor $\<12\>^8$, such polynomials always produce local matrix elements.

In particular, for $k=4$ --- i.e.~$D^8 R^n$ --- an example of a superamplitude can be written down explicitly. This requires a square bracket polynomial
 that is fully symmetric under exchanges of any two momenta and has each square spinor appearing precisely $4$ times. There are two obvious choices:\bea
 \frac{m_n(--++\dots+)}{\<12\>^8}
 =
 c_1 \big([12]^2 [23]^2 \cdots [n1]^2 + \text{perms}\big)
 +c_2 \big(([12] [34] \cdots [n-1,n])^4 + \text{perms}\big) \, .
\eea
The second term only exists if $n$ is even, but the first is valid for all $n$.\footnote{The first term was also identified in \cite{Kalloshtalk}.}  For $n=6$ the two terms are independent, and there are no other independent contributions.
For $n=4$ these two terms are linearly dependent through the Schouten identity.
Other structures become available as $k$ becomes sufficiently large; we will encounter this for $n=4$ in the following section.

\subsection{$D^{2k}R^4$ counterterms}
\lab{secD2kR4}

With the exception of $D^2 R^4$, which is excluded by kinematics, our analysis does not rule out candidate counterterms $D^{2k} R^4$. Instead our locality and symmetry constraints determine a unique superamplitude for $k\le5$, \ie $L \le 8$. For $L\ge9$ more than one structure is available. As explained, our method constructs
a symmetric function $m_4(--++)/\<12\>^8$, where  $m_4(--++)$ is local and has dimension $2k+8$ and the correct scaling weights. The expressions in terms of spinor brackets are converted to polynomials in $s$, $t$, and $u$. These are listed in Table \ref{tab4pt} for $0 \le k \le  7$, i.e.~$L=3,\dots, 9$.

One may compare these results to the $\alpha'$-expansion of the closed string 4-graviton Virasoro-Shapiro amplitude
\be
  \lab{VS4}
  M_4 ~=~ \frac{\G(1+\frac{\alpha'}{4} s)\G(1+\frac{\alpha'}{4} t)\G(1+\frac{\alpha'}{4} u)}
  {\G(1-\frac{\alpha'}{4} s)\G(1-\frac{\alpha'}{4} t)\G(1-\frac{\alpha'}{4} u)} ~M_4^\text{SUGRA tree} \, .
\ee
Since there are no free parameters in string theory, the matrix elements it determines for the operators $D^{2k} R^4$ are necessarily unique: they correspond to choosing the arbitrary constants in the field theory results to be specific combinations of $\zeta$-functions. These constants are listed in Table  \ref{tab4pt}.

The $\alpha'$ expansion of \reef{VS4} was studied in the papers \cite{
Green:2008uj}, and the first appearance of two independent structures in
$D^{12}R^4$ was noted \cite{Green:2008bf}. The expansion has also been studied recently in
\cite{Stieberger:2009rr}. Our analysis of polynomials in spinor brackets shows that these structures are the most general ones compatible with SUSY Ward identities and locality. In field theory the coefficients of these two structures are unfixed, while string theory selects a particular linear combination.

\begin{table}
\begin{eqnarray*}
\begin{array}{clcl}
\text{loop} & \text{counterterm}& f_k(s,t,u) &
 c_L~\text{in string theory (($\alpha'/4)^L$ suppressed)} \\[2mm]
3 & ~~~~R^4 & c_0 & ~~~~c_0 = 2 \zeta(3) \\[2mm]
4 & ~~~~D^2 R^4 &  \text{excluded} & ~~~~\text{absent} \\[2mm]
5 & ~~~~D^4 R^4 & c_2\, (s^2 + t^2 + u^2)& ~~~~c_2 = -\zeta(5) \\[2mm]
6 & ~~~~D^6 R^4 & c_3\, s\, t\, u & ~~~~c_3 = 2\zeta(3)^2 \\[2mm]
7 & ~~~~D^8 R^4 & c_4\, (s^2 + t^2 + u^2)^2  & ~~~~c_4 = \frac{1}{2}\zeta(7) \\[2mm]
8 & ~~~~D^{10} R^4 & c_{5}\, s\, t\, u\, (s^2 + t^2 + u^2)  & ~~~~c_{5} = 2\zeta(3)\zeta(5) \\[2mm]
9 & ~~~~D^{12} R^4 & c_{6}\, s^2\, t^2\, u^2 +  c_{6}' (s^3 t^3 + t^3 u^3 + u^3 s^3)  & ~~~~c_{6} =-\frac{4}{3} (\zeta(3)^3 + 5 \zeta(9))\,,~\,c_{6}' = 2\zeta(9)
\end{array}
\end{eqnarray*}
\vspace{-5mm}
\caption{{\small Matrix elements $m_4(--++) =
A_\text{tree} \times s\, t\, u  \times f_k(s,t,u)$ of $D^{2k} R^4$ for $0 \le k \le  6$. In $\cn=8$ supergravity these are matrix elements of candidate counterterms for loop levels $L=3,\dots, 9$.}}
\lab{tab4pt}
\end{table}


\setcounter{equation}{0}
\section{Candidate NMHV counterterms}
\lab{secNMHV}

Superamplitudes in the NMHV sector of $\cn=8$ supergravity are Grassmann polynomials of order 24 in the variables $\h_{ia}$.  To study potential counterterms  we use the superamplitude representations recently derived in \cite{efk4}. This construction guarantees that individual matrix elements, obtained 
by Grassmann differentiation, are related by the appropriate SUSY Ward 
identities.   Since the Ward identities are under control, we can proceed to study whether all NMHV matrix elements for a fixed number $n$ of external particles can be local functions, i.e.~polynomials, in the spinor brackets $\<ij\>,~[kl]$.

In this section we rule out $R^n$ and $D^2R^n$ NMHV counterterms by a shift argument similar to  that used at the MHV level in section~\ref{secMHV}. The NMHV bound is weaker than in  the MHV sector where we also ruled out independent $D^{4}R^n$ and $D^{6}R^n$ counterterms. 

\subsection{No $R^6$ and $D^2R^6$ NMHV counterterms}

\label{secR6D2R6}
The 6-point superamplitude of \cite{efk4}
 requires the input of 5 independent basis matrix elements, specifically the 6-graviton matrix element
$m_6(-+++--)$, and four other matrix elements in which the first two particles are pairs of gravitini, graviphotons, graviphotini, and scalars:
\bea
  \nonumber
   {\cal C}_6^{\rm NMHV} &=&
   \Big\{~~
   m_6(    -\, + \,+ + -\, -)\,\, X_{\,11111111\,}
   \,~+~ m_6(   \psi^{-}\psi^{+}  + + - -)\,
   X_{(11111112)}
   \\[1mm]  \nonumber
   &&~\,
   +m_6( v^-v^++ + - -)\,
   X_{(11111122)}
   ~+~m_6(\chi^{-}\chi^{+} + + - -)\,
   X_{(11111222)} \\[1mm]
   &&~\,
   +\frac{1}{2}m_6(\, \phi^{1234} \phi^{5678} ++  - -)\,
   X_{(11112222)} \Big\}
   ~~+~~ \text{(1 $\leftrightarrow$ 2)}\, .
   \lab{SGNMHV}
\eea
The polynomials $X_{ijklpquv} $  are the 
24th order SUSY invariant Grassmann polynomials
\begin{equation}\label{xpoly8}
\begin{split}
   X_{ijklpquv} ~&=~ \d^{(16)}\Bigl(\sum_i | i\> \eta_{ai}\Bigr)~
 \frac{\,\pdan_{i,n\dash 3,n\dash 2;1}\,\,\pdan_{j,n\dash 3,n\dash2;2}\,\cdots\,\pdan_{v,n\dash 3,n\dash 2;8}\,}{[n-3,n- 2]^8\<n-1,n\>^8 }\,\,,\\
\pdan_{ijk,a} &\equiv [ij] \h_{ka} +[jk] \h_{ia} +[ki] \h_{ja}\,.
\end{split}
\end{equation}
 The parentheses in the polynomials $X_{(\ldots)}$ in~(\ref{SGNMHV})  indicate symmetrization  in the positions of the labels $1,2$. For example, $X_{(11111112)}=X_{11111112}+X_{11111121}+\ldots$~.
Each polynomial  $X_{(\ldots)}$ in~(\ref{SGNMHV}) is SUSY and R-symmetry invariant.
It is the action of the $\tQ_a$ and $Q^a$ SUSY generators that permits us to `standardize' the basis, so that all basis matrix elements contain 2 positive and 2 negative helicitity gravitons~\cite{efk4}.  In \reef{SGNMHV}, these have been chosen as lines 3,4 and 5,6, respectively.   This means that all $X$-polynomials contain the factor  $1/([34]\<56\>)^8$.

If $D^{2k}R^6$ has a supersymmetrization, there is a corresponding 
superamplitude of the form \reef{SGNMHV} that generates its $6$-point 
matrix elements. Each basis matrix element must be a local expression of 
mass dimension $2(k+6)$, so the total number of angle and square spinors is
$\sum_i(a_i +s_i) = 4(k+6)$. The difference  is determined by the sum of the helicity weights, hence $\sum_i(a_i -s_i)= -2\sum_i h_i =  0$  for any basis element of \reef{SGNMHV}.  Thus each basis matrix element is a product of  $6+k$ angle and $6+k$ square brackets.\footnote{Although we need this information only for basis matrix elements, it is also correct for all other NMHV 6-point  matrix elements.}

Using a suitable complex shift, we now show that (when $k=0,1$) the potential pole factor $1/ \<56\>^8$ \emph{cannot cancel} 
in  the permuted $6$-graviton matrix element $m_6(--+++-)$  obtained from the superamplitude~(\ref{SGNMHV}). We project out $m_6(--+++-)$ from the superamplitude by applying the Grassmann derivatives associated with the negative helicity graviton lines,
\begin{equation}\label{permm6}
    m_6(--+++-)~=~\biggl[\,\prod_{i=1,2,6}\prod_{a=1}^8 \frac{\pa}{\pa\h_{ia}}\biggr]\,{\cal C}_6^{\rm NMHV}\,.
\end{equation}
All basis matrix elements in~(\ref{SGNMHV}) contribute. To simplify notation, we denote the basis matrix element in  \reef{SGNMHV}
whose  $X$-polynomial is labeled by $(8-j)$ 1's and $j$ 2's by $m^{(j)}$.   Thus  $m^{(0)} = m_6(-+++--),~ m^{(1)}=m_6(\psi^-\psi^+++--), ~m^{(6)} =m_6( v^+v^-++--)$, etc;  the last example is in the $1\lra 2$ exchanged part of \reef{SGNMHV}.
With some  attention one can see that the Grassmann differentiations in~(\ref{permm6}) produce a coefficient of  $\<26\>^{8-j}\<16\>^j/\<56\>^8$ for the basis matrix element $m^{(j)}$. Thus we obtain
\be  \lab{perm6}
m_6(--+++-)\,=\, \frac{1}{\<56\>^8} \sum_{j=0}^8 {8\choose j} \,\<26\>^{8-j}\<16\>^j\, m^{(j)}\,.
\ee
The eight angle brackets in the numerator come from derivatives
of the Grassmann $\delta^{(16)}$ in the $X$-polynomials~(\ref{xpoly8}). The factor $1/[34]^8$   in \reef{SGNMHV} cancels in \reef{perm6}  because differentiation
of the $m_{ijk,a}$ polynomials produces compensating factors in all terms.
The binomial coefficients appear because of the symmetrization of labels in the $X$-polynomials.

Consider now the effect  of a holomorphic
 $3$-line shift of angle spinors as in~(\ref{shift}), but acting on the spinors $|3\>$, $|4\>$, and $|5\>$.  Spinor brackets $\<q \,q'\>$ are invariant under this shift unless they involve at least one spinor from the set $|3\>$, $|4\>$, $|5\>$. Shifted brackets are linear in $z$.  The denominator of \reef{perm6} has an 8th order pole in $z$,  but the brackets $\<26\>$ and $\<12\>$ in the numerator do not shift. The only potential $z$ dependence in the numerator comes
from the $6+k$ spinor brackets in the basis matrix elements $m^{(j)}$.
The pole cannot cancel in any linear combination of basis elements if  they contain fewer than 8 shifted angle brackets.  Thus the counterterm is ruled out if $6+k<8$; hence for $ k= 0,\,1$\,.

One may suspect that there could be supersymmetric operators of dimension 12 (like $R^6$) or 14 (like $D^2R^6$) that do not have a leading pure-graviton component. $m_6(--+++-)$ vanishes in this case, and the above shift argument would be void. In section \ref{secnograv} we will show that no such operators exist.

The conclusion is that potential $R^6$ and  $D^{2}R^6$ NMHV level counterterms cannot exist. Our argument does not show whether $k=1$ is an actual upper bound for ruling out  candidate NMHV $D^{2k}R^6$ counterterms, or simply the place where the shift argument above breaks down.  We will show that it is the actual bound by constructing a candidate $D^4R^6$ counterterm in section \ref{secD4R6}.

\subsection{No $R^n$ and $D^2R^n$ NMHV counterterms for $n \ge 6$}
\label{secRnD2Rn}

 We now generalize the analysis of the previous section to all $n\geq6$.
The superamplitude of a $D^{2k}R^n$ NMHV counterterm takes the general form \cite{efk4}
\begin{equation}
    {\cal C}_n^{\rm NMHV} ~=~ \sum_j m^{(j)} X_{(j)}\,.
\end{equation}
The sum includes all $n$-point basis matrix elements $m^{(j)}$  of the form $m_n(\cdots ++--)$.
Here the last 4 particles are the standardized graviton states we have discussed, and 
the $\cdots$
 indicate states of  $n-4$ particles of the $\cn =8$ theory such that the overall configuration is NMHV.
The $X_{(j)}$ are symmetrizations of the  polynomial $X_{ijklpquv}$ defined in~(\ref{xpoly8}) with
$1\le i\le j\le\ldots \le v \le n-4$.
Dimensional analysis, little group scaling and locality determine that the $m^{(j)}$ are polynomials of order  $6+k$ in angle brackets and order $2n+k-6$ in square brackets.

To test locality, we compute a non-basis matrix element with three positive helicity graviton lines from this superamplitude. The result takes the form
\begin{equation}\lab{permn}
    m_n(\cdots + + + -)~=~\frac{1}{\<n\,\dash 1,n\>^8}\sum_j f_{(j)}\!\bigl(\<1n\>,\<2n\>,\ldots,\<n\dash4,n\>\bigr)\,m^{(j)}\,,
\end{equation}
where the $f_{(j)}$'s are some polynomials of total degree 8 in the indicated angle brackets. This polynomial arises from the differentiation of the Grassmann delta function in the $X_{(j)}$ and, crucially, it therefore does not contain any angle spinors from the positive helicity graviton lines $n-3$, $n-2$, and $n-1$.

We now  perform a holomorphic $3$-line shift~(\ref{shift}) of the three positive helicity lines $n-3$, $n-2$, and $n-1$. The shifted denominator has an 8th order pole in $z$. In the numerator, the $f_{(j)}$ are invariant under the shift because they all only depend on unshifted angle brackets. The $m^{(j)}$ contain $6+k$ angle brackets and are therefore at most $O(z^{6+k})$ under the shift. Again  the 8th order pole cannot be canceled for $k\leq1$.

As in the $n=6$ case, one may worry that there could be  NMHV operators whose matrix elements $m_n(\cdots + + + -)$ vanish for any assignment of the first $n-4$  external states $\cdots$. This would invalidate the above shift argument. In section \ref{secnograv} we will show that no such operators exist.

We conclude that there are no independent 
 supersymmetrizations of $R^n$ or $D^2R^n$ at the NMHV level for $n \ge 6$.


\setcounter{equation}{0}
\section{Matrix elements of a $D^4R^6$ NMHV operator}\label{secD4R6}

In this section, we construct an explicit permutation symmetric superamplitude for the NMHV operator $D^4R^6$ to demonstrate that it is not excluded. We use the representation \reef{SGNMHV} for the NMHV 6-point superamplitude, so the only input needed are the 5 independent basis matrix elements. These matrix elements must be local and have mass dimension 16. A direct approach to construct these matrix elements was unworkable for Mathematica, and instead we devised a method to construct the $\cn=8$ superamplitude from the product of $\cn=4$ gauge theory superamplitudes.\footnote{This is \emph{not} the same as the KLT relations \cite{Kawai:1985xq} for tree amplitudes.}
We describe the construction in section \ref{s:ggt} before turning to its practical implementation in sections \ref{s:D2F6} and \ref{s:D4R6}.

\subsection{Gravity from (gauge theory)$^2$}
\lab{s:ggt}

Let $\mathcal{F}$ and   $\mathcal{\tilde{F}}$  be
NMHV $n$-particle $\cn=4$ SYM theory superamplitudes. Suppose that they are $SU(4)_R$ invariant and are annihilated by the $\cn=4$ SUSY charges, $\tilde{Q}_A \mathcal{F} = Q^A \mathcal{F} =0$. As superamplitudes
for color-ordered amplitudes, $\mathcal{F}$ and  $\mathcal{\tilde{F}}$ have dihedral (cyclic and reflection) symmetry. However, we would like to use them in a gravity construction, so we impose full $S_n$ permutation symmetry and use the designation $\mathcal{F}^\text{sym}$ and $\mathcal{\tilde{F}}^\text{sym}$.

Now consider the product $\mathcal{C}^\text{NMHV} = \mathcal{F}^\text{sym}
\times \tilde{\mathcal{F}}^\text{sym}$ where the $SU(4)_R$-symmetry indices of $\mathcal{F}^\text{sym}$ are $A=1,2,3,4$ and those of $\tilde{\mathcal{F}}^\text{sym}$ are $\tilde{A}=5,6,7,8$. The $\cn=8$ SUSY charges split accordingly, hence $\mathcal{C}^\text{NMHV}$ is annihilated by the full set of $\cn=8$ SUSY generators. By construction, it is also permutation symmetric, so the only property it lacks is the full $SU(8)_R$ symmetry; it only has $SU(4)_R \times SU(4)_R$ due to the particular split of the eight $SU(8)_R$ indices.  However, if we sum over all
(8 choose 4)=70 embeddings of $SU(4) \times SU(4)$ into $SU(8)$, then the resulting Grassmann polynomial
\be
  \lab{CFF}
  \mathcal{C}^\text{NMHV} 
  = \sum_{70\;\text{embeddings}}
  (\pm )\,
  \mathcal{F}^\text{sym}
  \times \tilde{\mathcal{F}}^\text{sym}
\ee
can easily be shown to have the full $SU(8)_R$-symmetry.  Here, the $(\pm)$ sign  is the signature of the permutation that brings the embedding $(A,\tilde{A})$ into the canonical order $(1,\ldots,8)$.
We note that if $\Delta_1$ and $\Delta_2$ are the mass dimensions of the basis matrix elements of $\mathcal{F}^\text{sym}$ and $\tilde{\mathcal{F}}^\text{sym}$, then the mass dimension of the matrix elements generated by $\mathcal{C}^\text{NMHV}$ will be $\Delta_1 +\Delta_2$.

In our specific application, we want $\Delta_1 + \Delta_2 = 16$ for the NMHV matrix elements of $D^4 R^6$. Possible gauge theory operators are $F^6$, $D^2 F^6$ and $D^4 F^6$ whose matrix elements have dimension $6$, $8$ and $10$. We find that $(\mathcal{F}^\text{NMHV}_{F^6})^\text{sym}$ vanishes,\footnote{For example, helicity weights uniquely fix the dimension 6 basis matrix element $a_6(-+++-\,-)$ to be
$\<23\>\<34\>\<42\>[15][56][61]$, whose symmetrization in, for instance, 5 and 6 vanishes.
}
so the only possibility is
\be
  \lab{CFF2}
  \mathcal{C}^\text{NMHV}_{D^4 R^6} = \sum_{35\;\text{embeddings}}
  \big(\mathcal{F}^\text{NMHV}_{D^2F^6}\big)^\text{sym}
  \times \big(\tilde{\mathcal{F}}^\text{NMHV}_{D^2F^6}\big)^\text{sym} \, .
\ee
Here, we only need to sum over 35=(8 choose 4)/2 embeddings, because the same  superamplitude is 
used for both $SU(4)_R$ factors, and  the 70 original embeddings then match up pairwise.
In the following subsection we construct
$(\mathcal{F}^\text{NMHV}_{D^2F^6})^\text{sym}$
and in section \ref{s:D4R6} we extract the five basis matrix elements of
$\mathcal{C}^\text{NMHV}_{D^4 R^6}$ from  \reef{CFF2}.
As a consistency check, we have explicitly verified that the result for $\mathcal{C}^\text{NMHV}_{D^4 R^6}$ correctly reproduces all permuted basis matrix elements.

\subsection{$D^2F^6$ in $\cn=4$ SYM}
\lab{s:D2F6}

The superamplitude $({\cal F}_{D^2F^6}^{\rm NMHV})^\text{sym}$
has a basis expansion of the form  \cite{efk4}
\begin{equation}\label{d2f6}
\begin{split}
  ({\cal F}_{D^2F^6}^{\rm NMHV})^\text{sym}~&=~ a^{(0)}_\text{sym} \,\,X_{(1111)}+a^{(1)}_\text{sym} \,\,X_{(1112)}
+a^{(2)}_\text{sym} \,\,X_{(1122)} ~+~(1\leftrightarrow2)\,,\\[2mm]
\text{with}~~~~
 X_{ijkl} ~&\equiv~ \d^{(8)}\Bigl(\,\sum_{i=1}^6|i\>\eta_{ia}\Bigr)
 ~\frac{\,m_{i34;1}\,\,m_{j34,2}\,\,m_{k34,3}\,\,m_{l34,4}\,}{[34]^4\<56\>^4}\,.
\end{split}
\end{equation}
The 3 basis matrix elements involve the gluons (denoted by $+$ and $-$), gluinos ($A^a$ and $A^{abc}$), and scalars $A^{ab}$ of $\cn=4$ SYM theory. The superscripts are $SU(4)_R$ symmetry indices. In \reef{d2f6} we use the shorthand notation 
 $a^{(0)}_\text{sym}= a_6(-+++--)_\text{sym}$,  $a^{(1)}_\text{sym} = a_6(A^{123}A^{4}++--)_\text{sym}$, and $a^{(2)}_\text{sym} = a_6(A^{12}A^{34}++--)_\text{sym}$.

The construction of $a^{(0)}_\text{sym}$ proceeds as follows. We (let Mathematica) construct  all possible local angle/square bracket contractions of dimension 8 compatible with the scaling weights. We find 177 such local terms, but when we impose
bose/fermi symmetry of identical particles, only 9 terms survive. However, several of these are dependent through Schouten and momentum conservation. This leaves only 3 independent terms with the correct symmetries, and we express $a^{(0)}_\text{sym}$ as a general linear combination of those. A similar construction is carried out for $a^{(1)}_\text{sym}$ and $a^{(2)}_\text{sym}$, as summarized in Table \ref{tabD2F6}. With these three local basis matrix elements we now have an ansatz for the superamplitude
$({\cal F}_{D^2F^6}^{\rm NMHV})^\text{sym}$. Requiring that it has full permutation symmetry fixes all parameters in the basis elements. This gives the result 
\begin{equation}\label{basis}
    a^{(0)}_\text{sym}=a_6(-+++--)_\text{sym}=\sum_{m_i,p_i} [p_1 p_3]\<m_1 m_3\> [p_2 p_3] \<m_2 m_3\> \<m_1|m_2\!+\!m_3|p_1] \<m_2|p_1\!+\!p_3|p_2] \,.
\end{equation}
The sum on $m_i$ and $p_i$ is over all permutations of $1,5,6$ and $2,3,4$, respectively. The expressions for $a^{(1)}_\text{sym}$ and $a^{(2)}_\text{sym}$ are more complicated; they are given in appendix \ref{appD2F6}.

\begin{table}
\begin{eqnarray*}
\begin{array}{lclccccc}
&&& {\scriptstyle (a)} & {\scriptstyle (b)} & {\scriptstyle (c)} & {\scriptstyle (d)} \\
  && ~~~\hspace{2.2cm}\text{Impose:} & \text{little grp}  &
\text{bose/fermi} & {}^\text{~~~Schouten}_\text{\& mom.cons.} 
& \text{conj.} \\[1mm]
   a^{(0)}_\text{sym} &=& a_6(-+++--)_\text{sym}  & 177  & 9 & 3
& \text{even} \to 2  \\[1mm]
   a^{(1)}_\text{sym} &=& a_6(A^{123}A^{4}++--)_\text{sym} & 684  & 166 & 24 &
   ~~\;\text{odd} \to 14  \\[1mm]
   a^{(2)}_\text{sym} &=& a_6(A^{12}A^{34}++--)_\text{sym} & 1115 & 189 & 24 &
   ~~\text{even} \to 15
\end{array}
~~\raisebox{-5mm}{\Bigg\}~$\to$ 1}
~~\hspace{-12mm}\raisebox{6.5mm}{perm.~sym.}
\end{eqnarray*}
\vspace{-4mm}
\caption{{\small Construction of basis matrix elements $a^{(i)}_\text{sym}$ for $D^{2} F^6$. In the columns we list how many terms are left after the constraint in the column caption is imposed. In column $(a)$, we construct all local matrix elements with the correct little group scalings.
In column $(b)$, the bose/fermi  exchange symmetries imposed on $a^{(0)}_\text{sym}$ are the permutations  $\mathcal{P}(1,5,6)$ and $\mathcal{P}(2,3,4)$. For  $a^{(1)}_\text{sym}$ they are $\mathcal{P}(3,4)$ and $\mathcal{P}(5,6)$, which are also imposed on $a^{(2)}_\text{sym}$ in addition to  $\mathcal{P}(1,2)$.
Column $(c)$ lists how many of the terms in column $(b)$ are independent with respect to  Schouten and momentum conservation.
All terms selected by SUSY are conjugation even/odd as stated in column $(d)$. \emph{Conjugation}  odd (even) here means that the terms (do not) change sign when angle/square brackets are exchanged and $\{ 1 \lra 2, \, 3 \lra 5, \, 4 \lra 6 \}$. Column $(d)$ lists how many of the terms in column $(c)$ have this conjugation structure.
Permutation symmetry selects a unique linear combination of the terms given in column $(d)$. These then determine a unique result for the superamplitude $({\cal F}_{D^2F^6}^{\rm NMHV})^\text{sym}$  (up to an overall constant). Of the 2 conjugation-even terms for $a^{(0)}_\text{sym}$ only one actually contributes, namely the one given in \reef{basis}. The terms needed for $a^{(1)}_\text{sym}$ and $a^{(2)}_\text{sym}$ are listed in appendix \ref{appD2F6}.} }
\lab{tabD2F6}
\end{table}

As an example of the conditions arising from requiring permutation symmetry, consider the ``alternating helicity'' matrix element $a_6(-+-+-\,+)_\text{sym}$, whose external particle assignments are related to the basis matrix element $a_6(-+++-\,-)_\text{sym}$ by exchange of lines $3\leftrightarrow 6$. From the superamplitude we find
\bea
      &&a_6(-+-+-+)_\text{sym}~=\biggl[\,\prod_{i=1,3,5}\,\prod_{a=1}^4\frac{\partial}{\partial \eta_{ia}}\biggr]~({\cal F}_{D^2F^6}^{\rm NMHV})^\text{sym}\\[1ex]
      \nonumber
      && \hspace{3mm} =~
      \frac{\<5|1\!+\!3|4]^4}{\<56\>^4[34]^4}\,a^{(0)}_\text{sym} ~-~ 4 \,\frac{\<5|1\!+\!3|4]^3\<15\>[24]}{\<56\>^4[34]^4}\,a^{(1)}_\text{sym}
      ~~~~+~6\,\frac{\<5|1\!+\!3|4]^2\<15\>^2[24]^2}{\<56\>^4[34]^4}\,a^{(2)}_\text{sym}~+~~\cdots
\eea
Although not obvious, the sum on the right-hand side turns out to be exactly of the form~(\ref{basis}), but this time the sum over $m_i$ is over all permutations of $1,3,5$, and the sum  over $p_i$ is over all permutations of $2,4,6$. Therefore this non-basis matrix element with permuted external lines is simply given by the corresponding momentum permutation of the basis matrix element, as required (cf.~section \ref{sec:perm}).

In summary, we have found that  $({\cal F}_{D^2F^6}^{\rm NMHV})^\text{sym}$, with the basis elements  described above, is an $S_6$ permutation symmetric, SUSY and R-symmetry invariant superamplitude that produces local matrix elements associated with  a linearly supersymmetrized $D^2F^6$ operator of  $\cn=4$ SYM.
We now use it to construct the superamplitude for the local matrix elements of a linearly supersymmetrized $D^4 R^6$ operator  in $\cn=8$ supergravity.

\subsection{NMHV  $D^4 R^6$ counterterm}
\lab{s:D4R6}

We obtain the superamplitude $({\cal C}_{D^4R^6}^{\rm NMHV})^\text{sym}$ as a product of gauge theory factors $({\cal F}_{D^2F^6}^{\rm NMHV})^\text{sym}$ using \reef{CFF2}. The sum over the 35 different embeddings of $SU(4)_R \times SU(4)_R \subset SU(8)_R$ gives
\begin{equation}
\begin{split}
  {\cal C}_{D^4R^6}^{\rm NMHV}~=&~~
  m^{(0)} \,\,X_{(11111111)}+
  m^{(1)} \,\,X_{(11111112)}+
  m^{(2)} \,\,X_{(11111122)}+
  m^{(3)} \,\,X_{(11111222)}\\
  &+\frac{1}{2}\,m^{(4)} \,\,X_{(11112222)}
   ~+~(1\leftrightarrow2)\,,
\end{split}
\end{equation}
where
\begin{equation}
\lab{mFROMa}
\begin{split}
  m^{(0)}&~=~m_6(-+++--)
  ~=~35\, [a^{(0)}_\text{sym}]^2\,,\\[1ex]
  m^{(1)}&~=~m_6(A^{1234567}A^{8}++--)
  ~=~35\, a^{(0)}_\text{sym}\,a^{(1)}_\text{sym}\,,\\[1ex]
  m^{(2)}&~=~m_6(A^{123456}A^{78}++--)
  ~=~15\, a^{(0)}_\text{sym}\,a^{(2)}_\text{sym}+20 [a^{(1)}_\text{sym}]^2\,, \\[1ex]
  m^{(3)}&~=~m_6(A^{12345}A^{678}++--)
  ~=~5\,  a^{(0)}_\text{sym}\, a^{(3)}_\text{sym}
  +30\, a^{(1)}_\text{sym}\,a^{(2)}_\text{sym}\,,\\[1ex]
  m^{(4)}&~=~m_6(A^{1234}A^{5678}++--)
  ~=~a^{(0)}_\text{sym}\,a^{(4)}_\text{sym}+16 \,a^{(1)}_\text{sym}\,a^{(3)}_\text{sym}
  +18 [a^{(2)}_\text{sym}]^2\, .
\end{split}
\end{equation}
The $a^{(i)}_\text{sym}$ with $i=0,1,2$ were constructed in the previous section, while $a^{(3)}_\text{sym}=a^{(1)}_\text{sym}|_{1 \lra 2}$ and
$a^{(4)}_\text{sym}=a^{(0)}_\text{sym}|_{1 \lra 2}$.

The numerical coefficients in \reef{mFROMa} are combinatorial factors from the particular ways the gravity $X$-polynomials are assembled from products of the ones in gauge theory. For example, $X_{(11111111)}$ can only arise as $X_{(11111111)} =X_{(1111)}\tilde{X}_{(1111)}$\,, and in each of the 35 embeddings the coefficient is the same, namely $[a^{(0)}_\text{sym}]^2$. This explains why $m^{(0)}=35\, [a^{(0)}_\text{sym}]^2$. Less trivially, consider $m^{(2)}$. Its polynomial $X_{(11111122)}$ can arise in three different ways. In (6 choose 2)=15 of the 35 embeddings of $SU(4)_R \times SU(4)_R \subset SU(8)_R$, it will come from either
$X_{(1111)}\tilde{X}_{(1122)}$ or $X_{(1122)}\tilde{X}_{(1111)}$; these both have coefficient $a^{(0)}_\text{sym}\,a^{(2)}_\text{sym}$. In the remaining (6 choose 3)=20 cases it comes from $X_{(1112)}\tilde{X}_{(1112)}$ which has coefficient $[a^{(1)}_\text{sym}]^2$. This accounts for the coefficients 15 and 20 in  the third line of \reef{mFROMa}.

We have explicitly verified that ${\cal C}_{D^4R^6}^{\rm NMHV}$ defined in this way is a SUSY invariant, R-invariant, and permutation invariant superamplitude. All matrix elements obtained from it are local polynomials of dimension 16.  So the SUSY and locality requirements for the matrix elements of a $D^4R^6$ counterterm  are satisfied.

The product construction we have outlined provides one superamplitude with the properties of a  candidate  $D^4R^6$ NMHV counterterm. There may be other independent candidates. It turns out that the single soft scalar limit of the above basis element $m^{(4)}$ does not vanish, so the corresponding operator is not $E_{7,7}(\mathbb{R})$ invariant. However, if there  are  also other independent $D^4R^6$ structures available, there may be a linear combination that does exhibit the low energy theorems expected  of an $E_{7,7}(\mathbb{R})$-invariant counterterm.


\setcounter{equation}{0}
\section{Counterterms with vanishing pure-graviton matrix elements?}
\lab{secnograv}

Heretofore, the main focus of our work has been the supersymmetrization of 
gravitational operators $D^{2k}R^n$. At MHV level, all supersymmetric operators are of this form because any MHV operator must have a non-vanishing $n$-graviton matrix element $m_n(--+\cdots+)$. This is obvious from its superamplitude~(\ref{mhvsamp}). Beyond the MHV level, however, the situation is more subtle.  For example, could an $n$-scalar N$^K$MHV counterterm, schematically $D^{2k} \phi^n$, have a supersymmetrization that does not include a purely gravitational operator? Such an operator could for example ``live'' above the ``$R^n$ diagonal" in the chart of Fig.~\ref{tabsummary}; or it could hide as an independent operator on or below the diagonal. We study such operators in this section.

The manifestly supersymmetric N$^K$MHV superamplitude \cite{efk4} can be expressed in terms of basis matrix elements that all involve at least four gravitons $++ -\,-$. Therefore, any operator with an independent supersymmetrization must include a component of the schematic form $D^{2k} R^4\, \Phi^{n-4}$, where $\Phi^{n-4}$ denotes any $n-4$ fields of the theory. Such an operator has mass dimension $\Delta \ge 8$. As a consequence, no such independent operator exists at the 1- and 2-loop levels for which $\Delta=4$ and $6$, but a separate analysis is required for $L \ge 3$. We now address this point at the NMHV level.

In section~\ref{secNMHV} we ruled out independent supersymmetrizations of $R^n$ and $D^2R^n$ at the NMHV level. We can write the exclusion statement as a bound on the mass dimension $\Delta=2(n+k)$:
\begin{equation}\label{bounds}
      \nexists\text{ indep.~NMHV SUSY operators with $\Delta<2n+4$} \,.
\end{equation}
We will now prove that this bound not only governs putative supersymmetrizations of $D^{2k}R^n$, but holds in general for any NMHV operator.

Let us begin with the  simplest case, namely $n=6$. Recall that we proved the non-existence of supersymmetrizations of $R^6$ and $D^2R^6$ in section~\ref{secR6D2R6} by exposing a non-locality in the matrix element $m_n(--+++\,-)$ through a shift argument. This shift argument would become vacuous if the graviton matrix element $m_n(--+++\,-)$ vanished. We show now that if the pure graviton matrix elements vanish, then the entire $6$-point NMHV superamplitude must vanish.\footnote{This result is a direct consequence of the basis expansion \reef{SGNMHV}  and holds  for basis elements of any dimension, whether  local or non-local.} We use the representation \reef{perm6} for $m_6(--+++\,-)$, which follows from \reef{SGNMHV}, as well as the analogous representations for other inequivalent permutations of the 6 lines.  Each such equation expresses a particular permuted 6-graviton amplitude as a linear combination of
basis elements $m^{(j)}$, which are kept arbitrary. Now suppose that all permuted 6-graviton matrix elements, including $m^{(0)}$ and $m^{(8)}$, vanish.
The rank of the resulting linear system reveals that no non-trivial solution for $m^{(1)},m^{(2)},\dots, m^{(7)}$ exists.  Thus no independent supersymmetric 6-point counterterms with vanishing all-graviton matrix elements exists, and hence the bound~(\ref{bounds}) holds for $n=6$.

 A similar  result holds for $n >6$. The validity of the shift argument applied to \reef{permn} breaks down  if all matrix elements $m_n(\cdots+++\,-)$ vanish. The \,$\cdots$\, denote $n-4$ particles of the theory consistent with $m_n(\cdots+++\,-)$ being NMHV. As above, we can show that the entire NMHV superamplitude vanishes in this case. For $n=7,\ldots,12$, we verified this using
the same strategy as in the $n=6$ case, but applied to the linear system obtained from
\reef{permn} and its permutations. Again the result is valid for any NMHV superamplitude, independent of dimension and locality.
For $n>12$, all basis matrix elements necessarily
contain (at least) three positive and one negative helicity graviton \cite{efk4}, so the superamplitude vanishes trivially if the matrix elements in this class vanish.

We conclude that the bound~(\ref{bounds}) holds for general operators, not just for supersymmetrizations of $D^{2k}R^n$. While there can be (linearly) supersymmetric NMHV operators that have vanishing $n$-graviton matrix elements for $n>6$, none of them can possibly live above the $D^4R^n$ ``line'' in Fig.~\ref{tabsummary}.

We suspect, but have not proven, that a generalization of the above NMHV result holds at the N$^K$MHV level. We can assume $K\leq n/2-2$ because 
an N$^K$MHV $n$-point superamplitude with $K> n/2-2$ can be treated as anti-N$^{(n\dash 4\dash K)}$MHV. We suspect that, just as at the NMHV level, 
a non-vanishing N$^K$MHV superamplitude must have at least one non-vanishing matrix element of the form $m_n(\cdots +++ -)$. Here the \,$\cdots$ represent arbitrary $n-4$ states consistent with the N$^K$MHV level. If this is indeed the case, a holomorphic three-line shift \reef{shift} of lines $n\dash3$, $n\dash2$, $n\dash1$ on this matrix element reveals that it is non-local if the basis matrix elements of the N$^K$MHV superamplitude contain less than 8 angle brackets. The number of angle brackets in an N$^K$MHV basis matrix element is $\Delta/2-n+4+2K$. We are thus led to conjecture that
\begin{equation}\label{conj}
      \text{conjecture: }~\quad\nexists\,\text{ indep.~N$^K$MHV SUSY operators with $\Delta<2n+8-4K$ ~for~ $n>4$}\,.
\end{equation}
 Combining $K\leq n/2-2$ with the bound \reef{conj} it follows that no independent dimension $\Delta<16$ supersymmetric operator exists whose leading matrix element appears beyond $n=4$ points.
 Thus, if the conjecture~(\ref{conj}) holds, 
a UV finite 4-point amplitude implies finiteness of all higher point amplitudes at the same loop level for $L<7$.


\setcounter{equation}{0}
\section{Summary of candidate counterterms in $\cn=8$ supergravity}
\lab{secDisc}

In this paper we have introduced an efficient method to analyze potential counterterms in $\cn=8$ supergravity. It tests whether the matrix elements of a putative counterterm operator could have a supersymmetric completion. The input is gauge invariance, locality, supersymmetry and R-symmetry invariance, little group scalings and dimensional analysis. We have applied it at the MHV and NMHV level, and in each case excluded a set of operators as independent candidate counterterms.

We now summarize our work and place it in the context of results and arguments already given in the literature. It is well-known that pure supergravity amplitudes are finite at 1-loop \cite{'tHooft:1974bx,Grisaru:1976ua} and 2-loops [\citen{Grisaru:1976nn,vanNieuwenhuizen:1976vb},\,\citen{Deser:1977nt}]. Our analysis has shown that no higher-point (graviton or non-graviton) SUSY operators with mass dimension $\Delta < 8$ exist. 
For 3-loops and higher:

\begin{itemize}
\item 3-loops. Dimensional analysis allows only $R^4$ as a candidate pure gravity counterterm. Our analysis identifies  the familiar  unique matrix element that satisfies all supersymmetry and locality constraints. This is not surprising, since it is well-known that $R^4$ is compatible with linearized supersymmetry \cite{Deser:1977nt}. In particular, it arises as the leading $\alpha'$-correction to the closed string 4-graviton amplitude \cite{Gross:1986iv}.
Superspace constructions exist for this term \cite{Kallosh:1980fi}. However, the explicit demonstration \cite{Bern:2007hh} that the 4-point 3-loop
 amplitude is UV finite means that $R^4$ is not generated as a counterterm
in perturbative $\cn=8$ super\-gravity.

\item 4-loops. The potential counterterms are $D^2R^4$ and $R^5$. It is known, and it is reproduced in our analysis in section \ref{secR4}, that the 4-point matrix element of $D^2R^4$ is excluded \cite{Drummond:2003ex}. The non-existence of the  $D^2R^4$ counterterm is consistent with the explicit results \cite{Bern:2009kd} that the 4-point 4-loop amplitude is finite.
It was argued in  \cite{Drummond:2003ex,Kallosh:2009jb} that $R^5$ is absent.
In our analysis $R^5$ is excluded as an independent counterterm since its matrix elements cannot be both  local and satisfy the SUSY Ward identities.

\item 5-loops. Our analysis shows that no independent supersymmetrization of $D^2 R^5$ and $R^6$ exists.  According to \cite{Drummond:2003ex}, harmonic $(8,2,2)$ superspace allows $D^4 R^4$ while string theory arguments  \cite{Pierre} indicate that it is not generated. 

\item 6-loops. We have shown that no independent supersymmetrizations of $D^4 R^5$,  $D^2 R^6$ and $R^7$ exist.  $D^6 R^4$ can be constructed in harmonic $(8,1,1)$ superspace  \cite{Drummond:2003ex}, but limits of string theory indicate its absence \cite{Pierre}. 
\end{itemize}

Our analysis proves that at loop orders $L<7$ no independent supersymmetric MHV or  NMHV candidate counterterms exist for $n$-point amplitudes with $n > 4$.
We have conjectured in section \ref{secnograv} that this also holds for $L<7$ at any N$^K$MHV level.

\begin{itemize}
\item 7-loops. Howe and Lindstrom \cite{Howe:1980th} constructed a
linearized
superspace counterterm corre\-sponding to $D^8 R^4$, but noted that it did not respect the full $E_{7,7}(\mathbb{R})$ symmetry. It was proposed in \cite{Bossard:2009mn} that another superspace construction of a 7-loop $E_{7,7}(\mathbb{R})$-invariant counterterm might exist.
Recent string theory analyses \cite{Green:2010sp} zoom in on this loop level as a likely possible first divergence.

Our analysis shows that $D^8 R^4$ is the only candidate  7-loop counterterm at the MHV level. In particular, this means that $D^6 R^5$ is absent.
However, we identify $D^4 R^6$ as a candidate counterterm for 6-point NMHV amplitudes. In section \ref{secD4R6} we used a gauge theory trick to construct the needed basis matrix elements of the
 superamplitude for this counterterm. The single soft scalar limits of the resulting matrix elements do not vanish.  However, our gauge theory based construction may not have yielded the most general counterterm.  Thus we cannot make any statement
 whether a supersymmetrization of $D^4 R^6$ with $E_{7,7}(\mathbb{R})$ symmetry exists or not.

An independent $D^2R^7$ counterterm (MHV or NMHV) is ruled out by our analysis, but we have not excluded the possibility of a
(linearized) supersymmetrization of $R^8$, 
which would only contribute at the N$^2$MHV level. Thus a 7-loop divergence could appear in the 8-point N$^2$MHV amplitude even if lower-point amplitudes at the same loop-level are finite.

\item 8-loops. Independently, Kallosh \cite{Kallosh:1980fi} and Howe and Lindstrom \cite{Howe:1980th} constructed an 8-loop superspace counterterm $D^{10}R^4$ with full $E_{7,7}(\mathbb{R})$ invariance. We are not aware of any approach that rules it out. In its absence, our method has identified $D^8 R^5$ as a possible independent counterterm. Higher-point counterterms are excluded at the MHV level at this loop order, but not beyond.

\item 9-loops.
At loop orders $L<9$, our method
identified unique supersymmetric local matrix elements for the operators $D^{2k} R^4$. Up to an overall constant, these therefore agree with the $\alpha'$-expansion of the 4-graviton closed string tree amplitude (see details in section \ref{secD2kR4}). However, at the 9-loop level, field theory allows 2 independent local matrix elements of $D^{12} R^4$ that  satisfy the SUSY Ward identities.\footnote{Note that $D^{12}R^4$ occurs as the counterterm for the 2-loop divergence in $D=11$ \cite{Bern:1998ug}.} One particular linear combination is selected by the string amplitude \cite{Green:2008uj,Green:2008bf}.
\end{itemize}

We commented on $E_{7,7}(\mathbb{R})$ in the above summary.
$E_{7,7}(\mathbb{R})$ is a global symmetry of the equations of motion of the classical $\cn=8$ theory \cite{Cremmer:1978ds,Cremmer:1979up}, and it manifests itself in tree-level amplitudes as low-energy theorems for soft-scalar limits \cite{BEF,ArkaniHamed:2008gz,Kallosh:2008rr}. In a regularization scheme that preserves the $E_{7,7}(\mathbb{R})$ symmetry,\footnote{We are also grateful to J.~Maldacena for discussions of this and related points.}
 on-shell matrix elements of counterterms should obey low-energy theorems of spontaneously broken $E_{7,7}(\mathbb{R})$. One can use this as an additional criterion to rule out candidate counterterms. 

Our analysis is strictly 4-dimensional since it makes heavy use of the spinor-helicity formalism. The study of counterterms for super Yang-Mills theory and supergravity in dimensions $D>4$ also illuminates the situation in $D=4$.
Perhaps one could combine our method with the recent higher-dimensional spinor-helicity constructions \cite{Cheung:2009dc,Boels:2009bv} to address such questions.

It would be interesting to extend our method in $D=4$ to study supersymmetrizations of operators at the N$^2$MHV level and beyond. 
In particular, it would be interesting to prove our conjecture in section \ref{secnograv}
that the mass dimension $\Delta$ of independent N$^K$MHV SUSY operators is bound by $\Delta < 2n +8 - 4K$. As we explained, this would imply that the only counterterms that are available at $L<7$ are supersymmetrizations of $D^{2k} R^4$.

 \section*{Acknowledgements}
We are indebted to
Nima Arkani-Hamed,
Zvi Bern, 
Freddy Cachazo,
Simon Caron-Huot,
Lance Dixon, 
Paul Howe,
Renata Kallosh,
Juan Maldacena, 
Donal O'Connell,
Kelly Stelle, 
Pierre Vanhove,
and Edward Witten 
for their generous help and discussions.
We thank Michele Papucci and Prentice Bisbal for help with scripts to
run our Mathematica codes on the IAS aurora cluster.

HE is supported in part by the US Department of Energy under grant DE-FG02-95ER40899 and NSF grant PHY-0503584.
The research of DZF is supported by NSF grant
PHY-0600465 and by the US Department of Energy through cooperative research agreement DE-FG-0205FR41360.
MK is supported by the NSF grant PHY-9802484.

\appendix

\setcounter{equation}{0}
\section{Local matrix elements are polynomials in angle and square brackets}
\lab{appA}

The $n$-particle matrix element of a scalar operator such as $\int d^4x \,\pa^{2k}\phi^n$ (with an unspecified distribution of  derivatives) must be a polynomial in the external momenta $p_i,~i=1,\ldots n$. In the spinor-helicity formalism, the momentum $p_i$ is described by the spinor bilinear 
$|i\>[i|$, so these matrix elements are polynomials in angle and square brackets. The analogous property for $n$-point matrix elements of operators
such as  $\int d^4x \,\pa^{2k} F^n$ in gauge theory or $\int d^4x\, \pa^{2k} R^n$ in gravity is a little more subtle because external particles now come dressed with polarizations. If $\e^\m_\pm(i)$ is the polarization vector of a gluon, we can write the polarization tensor of the graviton as   $\e^{\m\n}_\pm(i)=\e^\m_\pm(i)\e^\n_\pm(i)$.
The  matrix elements are then polynomials in the scalar products
$p_i\cdot p_j,~p_i\cdot \e(j)$ or $\e(i)\cdot\e(j)$.

In the spinor helicity formalism, the polarization vectors $\e^\m_\pm$ are bispinors of the form:
\be \lab{polvecs}
{\rm -ve~ helicity}\quad \e_-(i) = \frac {|i\>[q_i|}{\sqrt{2}\,[i\,q_i]}\qquad\qquad  {\rm +ve ~helicity}\quad \e_+(i) = \frac {|i]\<q'_i|}{\sqrt{2}\,\<i\,q'_i\>}\,.
\ee
Because of gauge invariance, one can choose arbitrary spinors $|q_i]$ and $|q'_i\>$ for each external line,  subject only to the conditions  $[i q_i] \ne 0,~~\<i\,q'_i\>\ne 0.$   The complete amplitude is independent of the choice of the $|q_i],~|q'_i\>$.  Thus an individual term  in the matrix element, which comes from a specific Wick contraction,  may have spurious poles when the denominators of \reef{polvecs} vanish, but these poles must cancel in the full matrix element. Thus we reach the conclusion\footnote{This argument is not valid for $n=3$: special kinematics allow denominator terms, as is well known from the $3$-point Parke-Taylor formula.} that the matrix element must be a polynomial in the available spinor brackets $\<ij\>,~[kl]$.

\setcounter{equation}{0}
\section{Matrix elements of $D^2 F^6$}
\lab{appD2F6}

The basis matrix element $a^{(1)}_\text{sym} = a_6(A^{123} A^{4} ++ --)_\text{sym} $ can be written as a sum of 12 terms, each of which is manifestly conjugation-odd:
\bea
    a^{(1)}_\text{sym} ~=
    \sum_{p_{1,2} \in P(3,4)}
    ~\sum_{m_{1,2} \in P(5,6)}
    ~\sum_{I=1}^{12} ~B_I
\eea
with
\bea
  \nonumber
  B_1 &=&
  +\frac{3}{2} \<1 m_1 \> [2p_1] \<1 m_2\> [2 p_2] \<2 m_1\> [1 p_1]
   \,\<m_2 | 1 - 2 | p_2] \, ,\\
  \nonumber
  B_2 &=&
  -\frac{9}{4} \<1 m_1 \>^2 [2p_1]^2 \<2 m_2\> [1 p_2]
   \,\<m_2 | 1 - 2 | p_2] \, , \\
  \nonumber
  B_3 &=& - \frac{3}{2} \<1 m_1 \> [2p_2] \<m_1 m_2\> [p_1p_2]
  \,\big\{ \<1 m_1\>^2 [1 p_1] [1 m_1] - [2p_1]^2 \<2 m_1\> \< 2 p_1\> \big\} \, ,\\
  \nonumber
  B_4 &=&
  -\frac{9}{4} \<1 m_1 \> [2p_1] \<2 m_2\> [1 p_2] \<m_1 m_2\> [p_1p_2]
   \<1 | m_1 -p_ 1 | 2] \, ,\\
  \nonumber
  B_5 &=&
  +\frac{3}{4} \<1 m_2 \> [2p_2] \<2 m_1\> [1 p_1] \<m_1 m_2\> [p_1p_2]
   \<1 | m_1 -p_ 1 | 2] \, ,\\
  B_6 &=&
  -\frac{15}{16} \<m_1 m_2 \>^2 [p_1 p_2]^2
  \,\big\{  \<1 m_1\> \<1 m_2\> [1 m_1] [2 m_2]
      -   [2 p_1] [2 p_2]\<2 p_1\> \<1 p_2\>  \big\} \, ,\\
  \nonumber
  B_7 &=&
  -\frac{21}{16} \<m_1 m_2 \>^2 [p_1 p_2]^2
  \,\big\{  \<1 m_1\>^2  [1 m_1] [2 m_1]
      -   [2 p_1]^2 \<2 p_1\> \<1 p_1\>  \big\} \, ,\\
  \nonumber
  B_8 &=&
  -\frac{9}{8} \<m_1 m_2 \>^2 [p_1 p_2]^2
  \,\big\{  \<1 m_1\>\<2 m_2\>  [2 m_1] [2 m_2]
      -   [2 p_1][1 p_2] \<1 p_1\> \<1 p_2\>  \big\} \, ,\\
  \nonumber
  B_9 &=&
  +\frac{3}{4} \<1 m_2 \>^2 [2 p_2]^2 \<2 m_1 \> [1 p_1]
   \,\<m_1| p_2 - m_2 |p_1] \, ,\\
  \nonumber
  B_{10} &=&
  +\frac{3}{8} \<1 m_1 \> [2 p_1] \<m_1 m_2\> [p_1p_2]
   \<p_1 m_2 \> [m_1 p_2]
   \,\<m_1| 1-2 |p_1] \, ,\\
  \nonumber
  B_{11} &=&
  -\frac{3}{4} \<m_1 m_2 \>^2 [p_1 p_2]^2
  \<p_1 m_1 \> [m_1 p_1]    \,\<1| m_2-p_2 |2] \, ,\\
  \nonumber
  B_{12} &=&
  +\frac{3}{8}  \<1 m_1 \> [2 p_1] \<m_1 m_2\> [p_1p_2]
  \,\big\{  \<m_2 p_2\>[p_2 m_1]\<m_1 p_1\>  [p_1 p_2]
      -       [p_2 m_2]\<m_2 p_1\>  [p_1 m_1] \<m_1 m_2\>  \big\} \, .
\eea

The basis matrix element $a^{(2)}_\text{sym} = a_6(A^{12} A^{34} ++ --)_\text{sym} $ can be written as a sum of 11 terms, each of which is manifestly conjugation-even:
\bea
    a^{(2)}_\text{sym} ~= \sum_{s_{1,2} \in P(1,2)}~\sum_{p_{1,2} \in P(3,4)}
    ~\sum_{m_{1,2} \in P(5,6)}
    ~\sum_{I=1}^{11} ~C_I
\eea
with
\bea
 \nonumber
  C_1 &=& -\frac{7}{4}\,
  \<s_1 m_1\> [s_1 p_1] \<s_1 m_2\> [s_1 p_2]
  \<s_2 m_1\>  [s_2 p_1] \<s_2 m_2\> [s_2 p_2]
  \, ,\\ \nonumber
  C_2 &=& \<s_1 m_1\> [s_1 p_1] \<s_2 m_2\>  [s_2 p_2]
  \big\{\< s_1 m_2\> \<s_2 m_1\> [s_1 p_1] [s_2 p_2] +
   \<s_1 m_1\> \<s_2 m_2\> [s_1 p_2] [s_2 p_1]\big\} \, ,\\ \nonumber
  C_3 &=&  \frac{1}{2}\, \<s_1 m_1\>^2 \<s_2 m_2\>^2 [s_1 p_2]^2 [s_2 p_1]^2 \, ,\\
  \nonumber
  C_4 &=& 3\, \<m_1 m_2\>  [p_1 p_2] \<s_1 m_2\> [s_1 p_2]
  \<s_2 m_1\>  [s_2 p_1]
  \big\{ s_{s_1 m_1} + s_{s_1 p_1} \big\} \, , \\ \nonumber
  C_5 &=& \frac{7}{4}\, \<m_1 m_2\>^2 [p_1 p_2]^2
  \<s_1 | \,p_1 \, s_2 \, m_1 | s_1] \, ,\\ 
  C_6 &=& \frac{5}{8}\, \<m_1 m_2\>^2 [p_1 p_2]^2
  \big\{\<s_1 |  \, m_1 \, s_2 \, m_2| s_1 ] +
   \<s_1 | \, p_1\, s_2 \, p_2 | s_1] \big\}\, ,\\ \nonumber
   C_7 &=& - \<m_1 m_2\>^2 [p_1 p_2]^2
   \,s_{s_1 m_1}\, s_{s_1 p_1}
    \, ,\\ \nonumber
   C_8 &=& -\frac{1}{4}\,
   \<m_1 m_2\>^2 [p_1 p_2]^2
   \big\{ s_{s_1 m_1}\, s_{s_1 m_2} 
   + s_{s_1 p_1}\, s_{s_1 p_2}
   \big\} \, ,\\ \nonumber
   C_9 &=& - \frac{3}{4}\,
  \<m_1 m_2\> [p_1 p_2]
  \<p_1 m_2\>  [m_1 p_2] \<s_1 m_1\>  [s_1 p_1]
  \<s_2 m_1\> [s_2 p_1]\, ,\\ \nonumber
  C_{10} &=& -\frac{1}{8}\,
  \<m_1 m_2\>^2  [p_1 p_2]^2
  \<p_1 | m_1\,  p_2\,  m_2|  p_1]
  \, ,\\  \nonumber
  C_{11} &=& \frac{1}{4}\,
  \<m_1 m_2\>^2 [p_1 p_2]^2 \, s_{p_1 m_1} \, s_{p_2 m_2}
  \, .
\eea


\end{document}